\documentclass[12pt,reqno]{amsart}
\usepackage{graphicx}
\usepackage{amscd}
\usepackage{amsmath}
\usepackage{epsfig}
\usepackage{amsfonts}
\usepackage{amssymb}

\providecommand{\U}[1]{\protect\rule{.1in}{.1in}}
\providecommand{\U}[1]{\protect\rule{.1in}{.1in}}
\allowdisplaybreaks

\theoremstyle{plain}

\numberwithin{equation}{section}

\input{tcilatex}

\begin{document}
\title[The $\langle r^{p}\rangle $ for Harmonic Oscillator in $\boldsymbol{R}%
^{n}$]{Expectation Values $\langle r^{p}\rangle $ for Harmonic Oscillator in 
$\boldsymbol{R}^{n}$}
\author{Ricardo Cordero-Soto}
\address{Mathematical, Computational and Modeling Sciences Center, Arizona
State University, Tempe, AZ 85287--1804, U.S.A.}
\email{ricardojavier81@gmail.com}
\author{Sergei K. Suslov}
\address{School of Mathematical and Statistical Sciences and Mathematical,
Computational and Modeling Sciences Center, Arizona State University, Tempe,
AZ 85287--1804, U.S.A.}
\email{sks@asu.edu}
\urladdr{http://hahn.la.asu.edu/\symbol{126}suslov/index.html}
\date{\today }

\begin{abstract}
We evaluate the matrix elements $\langle r^{p}\rangle $ for the $n$%
-dimensional harmonic oscillator in terms of the dual Hahn polynomials and
derive a corresponding three-term recurrence relation and a Pasternack-type
reflection relation. A short review of similar results for nonrelativistic
hydrogen atom is also given.
\end{abstract}

\maketitle

\section{An Introduction}

The purpose of this note is to present a simple evaluation of the
expectation values $\langle r^{p}\rangle $ for the $n$-dimensional harmonic
oscillator in terms of the dual Hahn polynomials by direct integration.
Other methods of solving similar problems in elementary quantum mechanics
appeal to general principles and involve the Hellmann--Feynman theorem (see 
\cite{Balas84}, \cite{Balas90}, \cite{Balas00} and references given there,
textbooks \cite{Kram}, \cite{Mes}, \cite{Ba:Zel:Per}, \cite{Per:Zel}),
commutation relation (\cite{HeyAug93}, \cite{Ep:Ep}, \cite{Rob65}, \cite%
{Rob67}), and dynamical groups (\cite{Armstr71a}, \cite{Armstr71b}, \cite%
{Barut:Beker:Rador}, \cite{Barut:Bek:Rad}, \cite{Chac:Lev:Mosh76}, \cite%
{Ojha:Croth84}, \cite{Ques:Mosh71b}). Our approach allows to study these
expectation values with the help of the advanced theory of classical
polynomials \cite{An:As:Ro}, \cite{Ko:Sw}, and \cite{Ni:Su:Uv}. We recall
also that the first-order of the time-independent perturbation theory
equates the energy correction to the expectation value of the perturbing
potential. Thus expressions of the form $\langle r^{p}\rangle $ gain utmost
importance.\medskip

The plan of the paper is as follows. We review results on expectation values 
$\langle r^{p}\rangle $ for the nonrelativistic hydrogenlike wave functions
first and then extend them, in a similar fashion, to the case of $n$%
-dimensional harmonic oscillator. An attempt to collect the available
literature is made.

\section{Expectation Values $\langle r^{p}\rangle $ for Coulomb Problems}

The problem of evaluation of matrix elements $\langle r^{p}\rangle $ between
nonrelativistic bound-state hydrogenlike wave functions has a long history
in quantum mechanics. An incomplete list of references includes \cite%
{Andrae97}, \cite{Armstr71a}, \cite{Balas00}, \cite{Ba:Zel:Per}, \cite{Beker}%
, \cite{Be:Sal}, \cite{Blanch74}, \cite{Bock74}, \cite{Chac:Lev:Mosh76}, 
\cite{Ep:Ep}, \cite{HeyAug93}, \cite{Kram}, \cite{La:Lif}, \cite{Libo}, \cite%
{Mes}, \cite{More83}, \cite{Ojha:Croth84}, \cite{Past}, \cite{PastCH}, \cite%
{Per:Zel}, \cite{Shab91}, \cite{Sus:Trey}, \cite{Swain:Drake90}, \cite%
{VanVleck34}, and \cite{Waller} and references therein. Although different
methods were used in order to evaluate these matrix elements, one of
possible forms of the answer seems has been missing until recently. In Ref.~%
\cite{Sus:Trey} the mean values for states of definite energy%
\begin{equation}
\left\langle r^{p}\right\rangle =\dfrac{\dint_{\mathbf{R}^{3}}\left\vert
\psi _{nlm}\left( \mathbf{r}\right) \right\vert ^{2}\ r^{p}\ dv}{\dint_{%
\mathbf{R}^{3}}\left\vert \psi _{nlm}\left( \mathbf{r}\right) \right\vert
^{2}\ dv}=\dfrac{\dint_{0}^{\infty }R_{nl}^{2}\left( r\right) r^{p+2}\ dr}{%
\dint_{0}^{\infty }R_{nl}^{2}\left( r\right) r^{2}\ dr},\quad
dv=r^{2}drd\omega  \label{in1}
\end{equation}%
in the nonrelativistic Coulomb problem have been evaluated in terms of the
Chebyshev polynomials of a discrete variable $t_{k}\left( x,N\right)
=h_{k}^{\left( 0,\ 0\right) }\left( x,N\right) $ originally introduced in
Refs.~\cite{Chebyshev59} and \cite{Chebyshev64}. Their extensions, the
so-called Hahn polynomials, introduced also by P.~L.~Chebyshev \cite%
{Chebyshev75} and given by%
\begin{eqnarray}
h_{k}^{\left( \alpha ,\ \beta \right) }\left( x,N\right) &=&\left( -1\right)
^{k}\frac{\Gamma \left( N\right) \left( \beta +1\right) _{k}}{k!\;\Gamma
\left( N-k\right) }  \label{hahn} \\
&&\times \ _{3}F_{2}\left( 
\begin{array}{c}
-k\medskip ,\ \alpha +\beta +k+1,\ -x \\ 
\beta +1\medskip ,\quad 1-N%
\end{array}%
;\ 1\right) ,  \notag
\end{eqnarray}%
were rediscovered and generalized in the late 1940s by W.~Hahn. (We use the
standard definition of the generalized hypergeometric series throughout the
paper \cite{Ba}, \cite{Erd}.)\medskip

The end results have the following closed forms%
\begin{equation}
\left\langle r^{k-1}\right\rangle =\frac{1}{2n}\left( \frac{na_{0}}{2Z}%
\right) ^{k-1}t_{k}\left( n-l-1,-2l-1\right) ,  \label{in2}
\end{equation}%
when $k=0,1,2,...$ and%
\begin{equation}
\left\langle \frac{1}{r^{k+2}}\right\rangle =\frac{1}{2n}\left( \frac{2Z}{%
na_{0}}\right) ^{k+2}t_{k}\left( n-l-1,-2l-1\right) ,  \label{in3}
\end{equation}%
when $k=0,1,...,\;2l.$ Here $a_{0}=\hslash ^{2}/me^{2}$ is the Bohr radius.
Equations (\ref{in1})--(\ref{in2}) reflect the positivity of the matrix
elements under consideration \cite{Sus:Trey}.\medskip

The ease of handling of these matrix elements for the discrete levels is
greatly increased if use is made of the known properties of these classical
polynomials of Chebyshev \cite{Erd}, \cite{Ko:Sw}, \cite{Ni:Su:Uv}, \cite%
{Ni:Uv} and \cite{Chebyshev59}, \cite{Chebyshev64}, \cite{Chebyshev75}. The
direct consequences of these relations are an inversion\ relation:%
\begin{equation}
\left\langle \frac{1}{r^{k+2}}\right\rangle =\left( \frac{2Z}{na_{0}}\right)
^{2k+1}\frac{\left( 2l-k\right) !}{\left( 2l+k+1\right) !}\ \left\langle
r^{k-1}\right\rangle  \label{in4}
\end{equation}%
with $0\leq k\leq 2l$ and the three-term recurrence relation:%
\begin{equation}
\left\langle r^{k}\right\rangle =\frac{2n\left( 2k+1\right) }{k+1}\left( 
\frac{na_{0}}{2Z}\right) \left\langle r^{k-1}\right\rangle -\frac{k\left(
\left( 2l+1\right) ^{2}-k^{2}\right) }{k+1}\left( \frac{na_{0}}{2Z}\right)
^{2}\left\langle r^{k-2}\right\rangle  \label{in5}
\end{equation}%
with the initial conditions%
\begin{equation}
\left\langle \frac{1}{r}\right\rangle =\frac{Z}{a_{0}n^{2}},\qquad
\left\langle 1\right\rangle =1,  \label{in6}
\end{equation}%
which is convenient for evaluation of the mean values $\left\langle
r^{k}\right\rangle $ for $k\geq 1.$\medskip

The recurrence relation (\ref{in5}) was originally found by Kramers and
Pasternack in the late 1930s \cite{Kram}, \cite{Past}, and \cite{PastCH}.
The inversion relation (\ref{in4}), which is also due to Pasternack, has
been rediscovered many years later \cite{Bock74}, \cite{More83} (see also 
\cite{HeyJan93} and \cite{HeyAug93} for historical comments).
Generalizations of (\ref{in4})--(\ref{in5}) for off-diagonal matrix elements
are discussed in Refs.~\cite{Blanch74}, \cite{Ojha:Croth84}, \cite{HeyAug93}%
, \cite{Shab91}, and \cite{Swain:Drake90}. The properties of the
hydrogenlike \ radial matrix elements are considered from a
group-theoretical viewpoint in Refs.~\cite{Armstr71a}, \cite{Chac:Lev:Mosh76}%
, and \cite{Ojha:Croth84}. Extensions to the relativistic case are given in 
\cite{Andrae97}, \cite{Davis}, \cite{Suslov}, and \cite{Sus:Trey} (see also
references therein).\medskip

In a retrospect, Pasternack's papers \cite{Past}, \cite{PastCH} had paved
the road to the discovery of the continuous Hahn polynomials in the mid
1980s (see \cite{KoelinkCH}, \cite{Sus:Trey} and references therein).

\section{Evaluation of $\langle r^{p}\rangle $ for Harmonic Oscillator in $%
\boldsymbol{R}^{n}$}

The stationary Schr\"{o}dinger equation for the $n$-dimensional harmonic
oscillator%
\begin{equation}
H\Psi =E\Psi ,\qquad H=\frac{1}{2}\sum_{s=1}^{n}\left( -\frac{\partial ^{2}}{%
\partial x_{s}^{2}}+x_{s}^{2}\right)  \label{h0}
\end{equation}%
can be solved by separation of the variables in hyperspherical coordinates.
The normalized wave functions have the form%
\begin{equation}
\Psi \left( \boldsymbol{x}\right) =\Psi _{NK\nu }\left( r,\Omega \right)
=Y_{K\nu }\left( \Omega \right) \ R_{NK}\left( r\right) ,  \label{h1}
\end{equation}%
where $Y_{K\nu }\left( \Omega \right) $ are the hyperspherical harmonics
associated with a binary tree $T,$ the integer number $K$ corresponds to the
constant of separation of the variables at the root node of $T$ and $\nu
=\left\{ l_{1},l_{2},...\ ,l_{p}\right\} $ is the set of all other
subscripts corresponding to the remaining vertexes of the binary tree $T$
(see \cite{Ni:Su:Uv}, \cite{Smir:Shit}, \cite{Smor}, and \cite{Vil} for a
graphical approach of Vilenkin, Kuznetsov and Smorodinski\u{\i} to the
theory of spherical harmonics).\medskip

The radial functions are given by%
\begin{equation}
R_{NK}\left( r\right) =\sqrt{\frac{2\left( \left( N-K\right) /2\right) !}{%
\Gamma \left( \left( N+K+n\right) /2\right) }}\ \exp \left( -r^{2}/2\right)
\ r^{K}\ L_{\left( N-K\right) /2}^{K+n/2-1}\left( r^{2}\right) ,  \label{h2}
\end{equation}%
where $L_{k}^{\alpha }\left( \xi \right) $ are the Laguerre polynomials \cite%
{Ni:Uv} and the corresponding energy levels are equal to%
\begin{equation}
E=E_{N}=N+n/2,\qquad \left( N-K\right) /2=k=0,1,2,...\ .  \label{h3}
\end{equation}%
(See also Refs.~\cite{Me:Co:Su}, \cite{Mosh:Ques71a}, \cite{Ni:Su:Uv}, \cite%
{Ques:Mosh71b}, and \cite{Smir:Shit} for group theoretical properties of the 
$n$-dimensional harmonic oscillator wave functions.)\medskip

The expectation values under consideration%
\begin{eqnarray}
\langle r^{p}\rangle &=&\int_{\boldsymbol{R}^{n}}\Psi _{NK\nu }^{\ast
}\left( r,\Omega \right) r^{p}\Psi _{NK\nu }\left( r,\Omega \right) \ dv
\label{h4} \\
&=&\int_{0}^{\infty }r^{p+n-1}R_{NK}^{2}\left( r\right) \ dr,  \notag
\end{eqnarray}%
we use orthonormality of the hyperspherical harmonics, are derived with the
aid of the integral \cite{Ojha:Croth84}, \cite{Sus:Trey}:%
\begin{align}
J_{nms}^{\alpha \beta }& =\int_{0}^{\infty }e^{-x}x^{\alpha +s}\
L_{n}^{\alpha }\left( x\right) L_{m}^{\beta }\left( x\right) \ dx  \label{h5}
\\
& =\left( -1\right) ^{n-m}\frac{\Gamma \left( \alpha +s+1\right) \Gamma
\left( \beta +m+1\right) \Gamma \left( s+1\right) }{m!\left( n-m\right)
!\;\Gamma \left( \beta +1\right) \Gamma \left( s-n+m+1\right) }  \notag \\
& \quad \times \ ~_{3}F_{2}\left( 
\begin{array}{c}
-m,\ s+1,\ \beta -\alpha -s\medskip \\ 
\beta +1,\quad n-m+1%
\end{array}%
\right) ,\quad n\geq m,  \notag
\end{align}%
where parameter $s$ may take some integer values. Similar integrals have
been discussed in Refs.~\cite{Andrae97}, \cite{Davis}, \cite{Doung}, and 
\cite{La:Lif}. A simple evaluation of this integral is given in Ref.~\cite%
{Sus:Trey}.\medskip

Indeed,%
\begin{equation}
\langle r^{p}\rangle =\frac{\left( \left( N-K\right) /2\right) !}{\Gamma
\left( \left( N+K+n\right) /2\right) }\ \int_{0}^{\infty }e^{-\xi }\xi
^{p/2+K+n/2-1}\ \left( L_{\left( N-K\right) /2}^{K+n/2-1}\left( \xi \right)
\right) ^{2}\ d\xi ,  \label{h5a}
\end{equation}%
where we replace $r^{2}=\xi .$ As a result,%
\begin{equation}
\langle r^{p}\rangle =\frac{\Gamma \left( K+\left( n+p\right) /2\right) }{%
\Gamma \left( K+n/2\right) }\ _{3}F_{2}\left( 
\begin{array}{c}
\left( K-N\right) /2\medskip ,\ p/2+1,\ -p/2 \\ 
K+n/2\medskip ,\quad 1%
\end{array}%
;\ 1\right) ,  \label{h6}
\end{equation}%
provided that $p+n+2K>0.$ Connection with the dual Hahn polynomials given by 
\cite{Ni:Su:Uv}%
\begin{eqnarray}
w_{m}^{\left( c\right) }\left( s\left( s+1\right) ,a,b\right) &=&\frac{%
\left( 1+a-b\right) _{m}\left( 1+a+c\right) _{m}}{m!}  \label{dhahn} \\
&&\ \times \ _{3}F_{2}\left( 
\begin{array}{c}
-m\medskip ,\ a-s,\ a+s+1 \\ 
1+a-b\medskip ,\quad 1+a+c%
\end{array}%
;\ 1\right)  \notag
\end{eqnarray}%
is as follows%
\begin{equation}
\langle r^{p}\rangle =\frac{\Gamma \left( K+\left( n+p\right) /2\right) }{%
\Gamma \left( \left( N+K+n\right) /2\right) }\ w_{\left( N-K\right)
/2\medskip }^{\left( 0\right) }\left( \frac{p}{2}\left( \frac{p}{2}+1\right)
,0,1-K-\frac{n}{2}\right) .  \label{h7}
\end{equation}%
Thus the direct calculation of $\langle r^{p}\rangle $ for each admissible $%
p $ involves a well-known special function originally introduced in Ref.~%
\cite{Karlin:McGregor61}. Basic properties of the dual Hahn polynomials are
discussed in \cite{Ni:Su:Uv}. We shall use one of them in the next section.
See also Refs.\ \cite{Chac:Lev:Mosh76}, \cite{Ques:Mosh71b}, and \cite%
{Armstr71b} for group-theoretical methods of evaluation of similar matrix
elements. We shall elaborate on the group-theoretical meaning of Eq.~(\ref%
{h6}) later.

\section{Three Term Recurrence Relation}

The difference equation for the dual Hahn polynomials has the form%
\begin{equation}
\sigma \left( s\right) \frac{\Delta }{\nabla x_{1}\left( s\right) }\left( 
\frac{\nabla y\left( s\right) }{\nabla x\left( s\right) }\right) +\tau
\left( s\right) \frac{\Delta y\left( s\right) }{\Delta x\left( s\right) }%
+\lambda _{m}y\left( s\right) =0,  \label{rr01}
\end{equation}%
where $\Delta f\left( s\right) =\nabla f\left( s+1\right) =f\left(
s+1\right) -f\left( s\right) ,$ $x\left( s\right) =s\left( s+1\right) ,$ $%
x_{1}\left( s\right) =x\left( s+1/2\right) ,$ and%
\begin{eqnarray}
&&\sigma \left( s\right) =\left( s-a\right) \left( s+b\right) \left(
s-c\right) ,  \label{rr02} \\
&&\sigma \left( s\right) +\tau \left( s\right) \nabla x_{1}\left( s\right)
=\sigma \left( -s-1\right)  \label{rr03} \\
&&\qquad =\left( a+s+1\right) \left( b-s-1\right) \left( c+s+1\right) , 
\notag \\
&&\lambda _{m}=m.  \label{rr04}
\end{eqnarray}%
It can be rewritten as a recurrence relation%
\begin{eqnarray}
&&\sigma \left( -s-1\right) \nabla x\left( s\right) y\left( s+1\right)
+\sigma \left( s\right) \Delta x\left( s\right) y\left( s-1\right)
\label{rr05} \\
&&\quad +\left( \lambda _{m}\Delta x\left( s\right) \nabla x\left( s\right)
\nabla x_{1}\left( s\right) -\sigma \left( -s-1\right) \nabla x\left(
s\right) -\sigma \left( s\right) \Delta x\left( s\right) \right) y\left(
s\right) =0.  \notag
\end{eqnarray}%
See \cite{Karlin:McGregor61}, \cite{Ko:Sw} and \cite{Ni:Su:Uv} for more
details on the properties of the dual Hahn polynomials.\medskip

Equations (\ref{h7}) and (\ref{rr05}) results in the three-term recurrence
relation%
\begin{eqnarray}
\left( p+2\right) \langle r^{p+2}\rangle &=&\left( p+1\right) \left(
2N+n\right) \langle r^{p}\rangle  \label{rr1} \\
&&+p\left( \frac{p^{2}-n^{2}}{4}-\left( K-1\right) \left( K+p-1\right)
\right) \langle r^{p-2}\rangle  \notag
\end{eqnarray}%
for the matrix elements. Such recurrence relation requires two initial
results%
\begin{equation}
\left\langle 1\right\rangle =1,\qquad \left\langle \frac{1}{r^{2}}%
\right\rangle =\frac{1}{K+n/2-1}  \label{rr2}
\end{equation}%
in order to generate all expectation values.\medskip

The cases $n=1,$ $2,$ $3$ are well-known and usually discussed with the help
of the hypervirial theorems (see, for example, \cite{Armstr71b}, \cite%
{Ba:Zel:Per}, \cite{Ep:Ep}, \cite{Balas90}, \cite{Per:Zel}, and references
therein.) Our approach emphasizes the simple fact that the matrix elements
recurrence relations (\ref{in5}) and (\ref{rr1}) for the hydrogen atom and
for the $n$-dimensional harmonic oscillator, respectively, have the same
\textquotedblleft special functions\textquotedblright\ nature, namely, they
occur due to the difference equation for the corresponding dual Hahn
polynomials. We leave further details to the reader.

\section{A Pasternack-Type Inversion Property}

From equation (\ref{h6}) one gets\ 
\begin{equation}
\langle r^{-p-2}\rangle =\frac{\Gamma \left( K+\left( n-p\right) /2-1\right) 
}{\Gamma \left( K+\left( n+p\right) /2\right) }\ \langle r^{p}\rangle
\label{ref1}
\end{equation}%
(for all the convergent integrals) in the spirit of Pasternack's inversion
relation (\ref{in4}) valid in the case of the hydrogen atom. It simply
reflects an obvious fact that the argument $x\left( p\right) =\left(
p/2+1\right) p/2$ of the dual Hahn polynomials in (\ref{h7}) obeys a natural
symmetry $x\left( -p-2\right) =x\left( p\right) $ on the corresponding
quadratic grid. We were unable to find relations (\ref{h6}), (\ref{h7}), and
(\ref{ref1}) in the available literature.

\section{A Conclusion}

Special functions appear in numerous applications of mathematical and
physical sciences and their basic knowledge is a necessity for any
theoretical physicist. As Dick Askey mentioned once in his talk on
mathematical education, \textquotedblleft We should not lie to our students,
but we do not have to tell them the whole truth.\textquotedblright\ The same
is true about teaching elementary topics in quantum mechanics. As we can
conclude from this short note, even if \textquotedblleft one is able to
evaluate average values by appeal to general principles rather that by
direct integration \cite{Ep:Ep}\textquotedblright , important relations with
the well-known special functions will be out of the picture due to the
limiting mathematical tools. Every instructor has to find a way how to
resolve this contradiction.\medskip 

\noindent \textbf{Acknowledgment.\/} This paper is written as a part of the
summer 2009 program on analysis of Mathematical and Theoretical Biology
Institute (MTBI) and Mathematical, Computational and Modeling Sciences
Center (MCMSC) at Arizona State University. The MTBI/SUMS Summer
Undergraduate Research Program is supported by The National Science
Foundation (DMS-0502349), The National Security Agency (DOD-H982300710096),
The Sloan Foundation, and Arizona State University. We thank Professor
Carlos Castillo-Ch\'{a}vez for support, valuable discussions and
encouragement. One of the authors (RCS) is supported by the following
National Science Foundation programs: Louis Stokes Alliances for Minority
Participation (LSAMP): NSF Cooperative Agreement No. HRD-0602425 (WAESO
LSAMP Phase IV); Alliances for Graduate Education and the Professoriate
(AGEP): NSF Cooperative Agreement No. HRD-0450137 (MGE@MSA AGEP Phase II).

\end{document}